\documentclass{article}
\usepackage{spconf,amsmath,graphicx}
\usepackage{amssymb}
\usepackage{amsmath}
\usepackage{bm}
\usepackage{hyperref}

\title{Parameter Efficient Transfer Learning \\for Various Speech Processing Tasks}
\name{Shinta Otake, Rei Kawakami, Nakamasa Inoue}
\address{Tokyo Institute of Technology}

\begin{document}
\maketitle

\makeatletter
\newcommand{\figcaption}[1]{\def\@captype{figure}\caption{#1}}
\newcommand{\tblcaption}[1]{\def\@captype{table}\caption{#1}}
\makeatother

\def\figminipage{
\begin{figure*}[t]
\def\@captype{table}
\begin{minipage}[]{.61\textwidth}
    \centering
    \tblcaption{{\bf Comparison of L-adapter configurations.} The best result is marked in bold. The top three results are marked with underline. }
    \vspace{10pt}
    \begin{tabular}{l|c|c|c|c|c}
    Config. & \# Params & ASV & ER & ASR & IC \\
    \hline
    Weight & 12 & $5.65^{\pm 0.25}$ & $28.9^{\pm 0.52}$ & $23.8^{\pm 0.13}$ & $0.73^{\pm 0.08}$ \\
    LN & 0.02M & $4.76^{\pm 0.07}$ & $25.8^{\pm 0.44}$ & $21.8^{\pm0.17}$ & $0.89^{\pm 0.08}$ \\
    Act+LN & 0.02M & $5.25^{\pm 0.23}$ & $26.9^{\pm 0.44}$ & $21.7^{\pm 0.29}$ & $0.73^{\pm0.08}$ \\
    FC & 4.72M & $3.71^{\pm 0.07}$ & $24.2^{\pm 0.31}$ & $16.8^{\pm 0.05}$ & $0.73^{\pm 0.11}$ \\
    FC+Act & 4.72M & $4.22^{\pm 0.09}$ & $\underline{22.8}^{\pm 0.55}$ & $\underline{10.2}^{\pm 0.05}$ & $0.67^{\pm 0.03}$ \\
    FC+LN & 4.74M & $\underline{\bm{2.73}}^{\pm 0.05}$ & $23.3^{\pm 0.51}$ & $13.5^{\pm 0.13}$ & $\underline{\bm{0.32}}^{\pm 0.03}$ \\\hline
    Base & 4.74M & $\underline{2.74}^{\pm 0.09}$ & $\underline{21.1}^{\pm 0.52}$ &  $\underline{\bm{9.50}}^{\pm 0.08}$ & $\underline{0.33}^{\pm 0.04}$ \\
    Skip & 4.75M & $\underline{2.78}^{\pm 0.06}$ & $\underline{\bm{20.0}}^{\pm 0.28}$ & $\underline{10.4}^{\pm0.08}$ & $\underline{0.42}^{\pm 0.09}$\\
    \end{tabular}
    \label{result1}
  \end{minipage}
  \hfill
  \begin{minipage}[]{.38\textwidth}
    \centering
    \includegraphics[width=7cm]{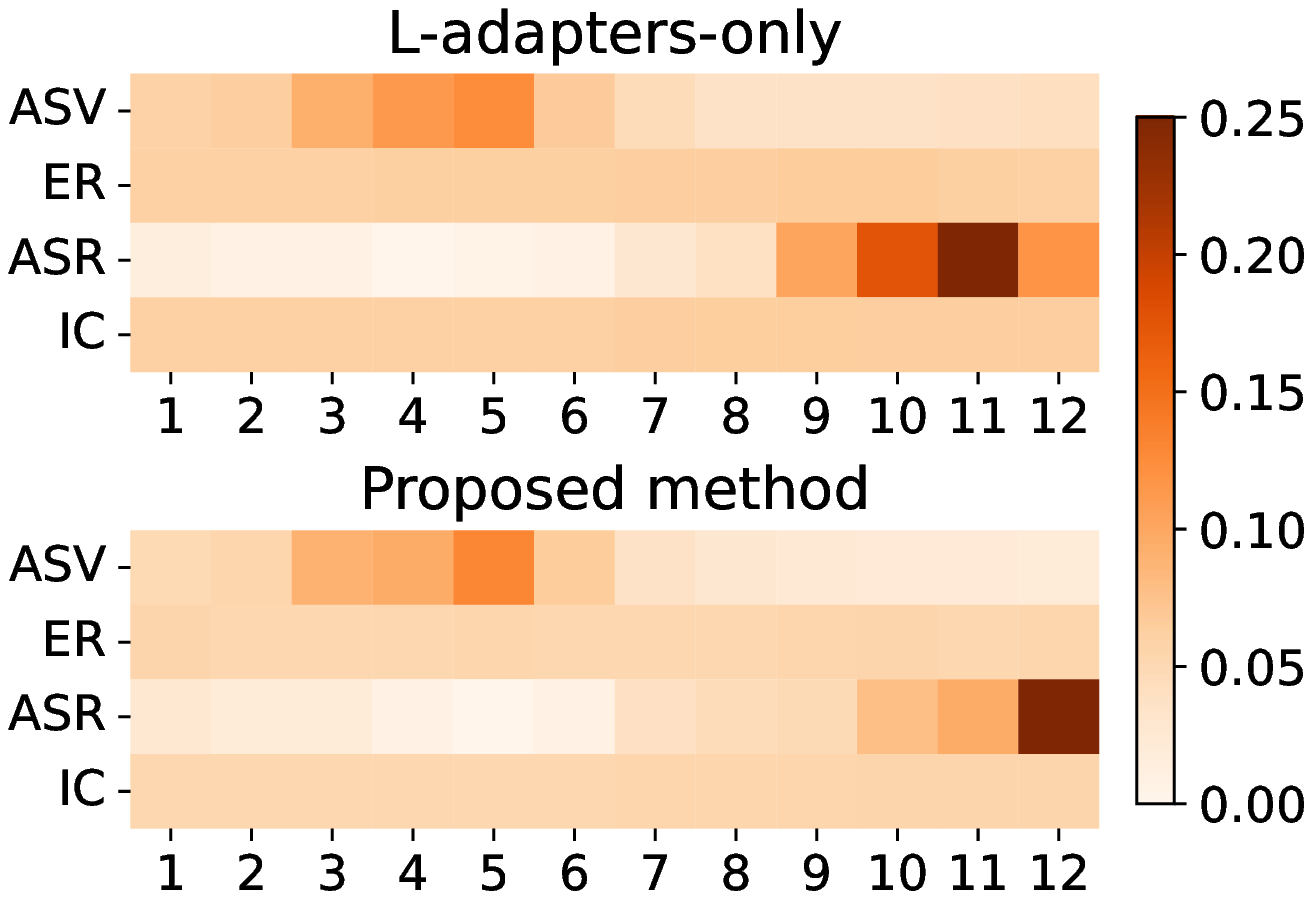}
    \vspace{-20pt}
    \caption{
    {\bf Layer weight analysis.}}
    \label{fig:layer-weight-analysis}
  \end{minipage}
\end{figure*}
}

\def\figlayeradapter{
\begin{figure}[t]
  \centering
  \includegraphics[width=9.1cm]{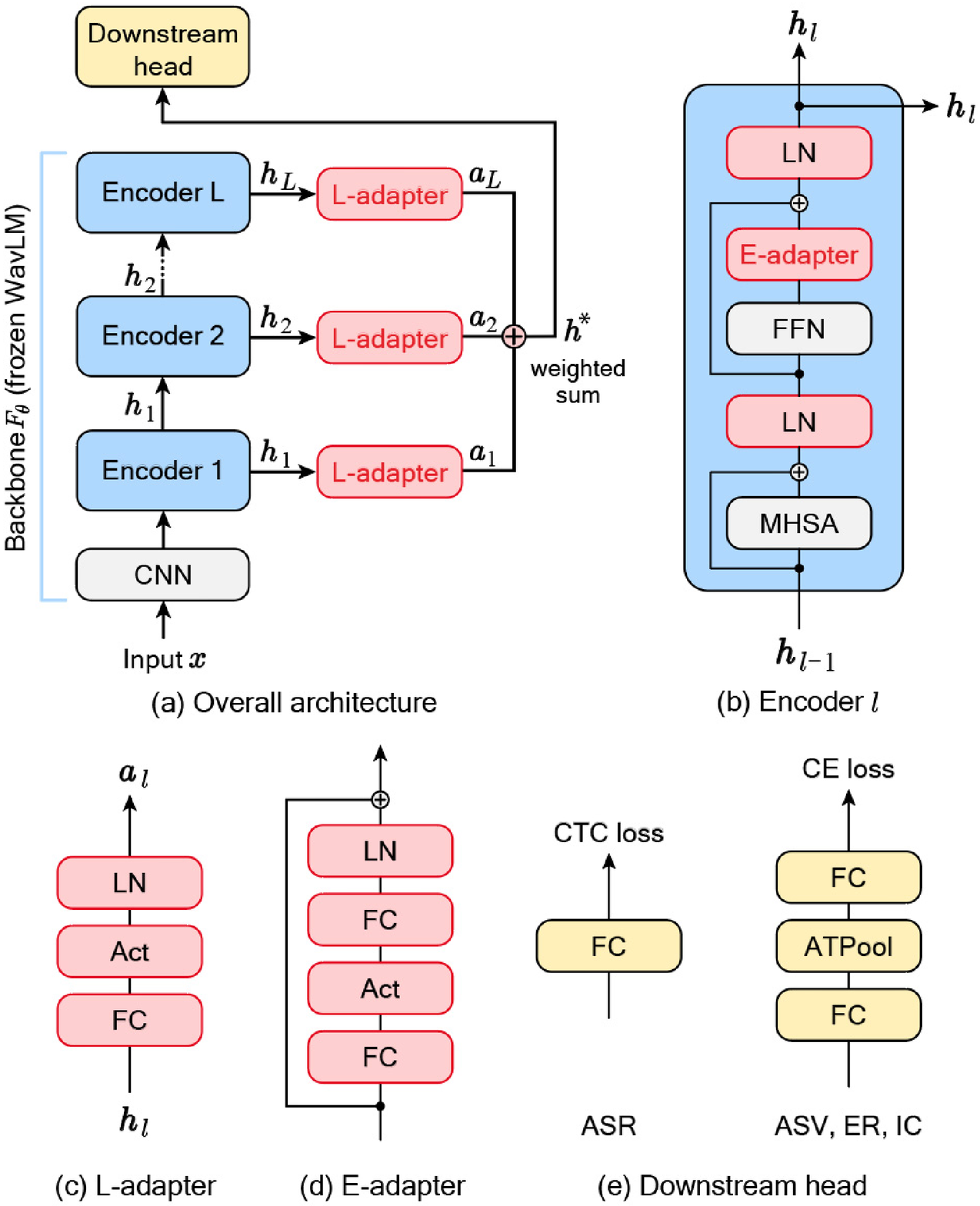}
  \vspace{-15pt}
\caption{
{\bf Proposed adapter architecture.}
}
\label{figlayeradapter}
\end{figure}
}

\def\figresults{
\begin{figure*}[t]
  \centering
  \includegraphics[width=21.0cm, height=7.0cm]{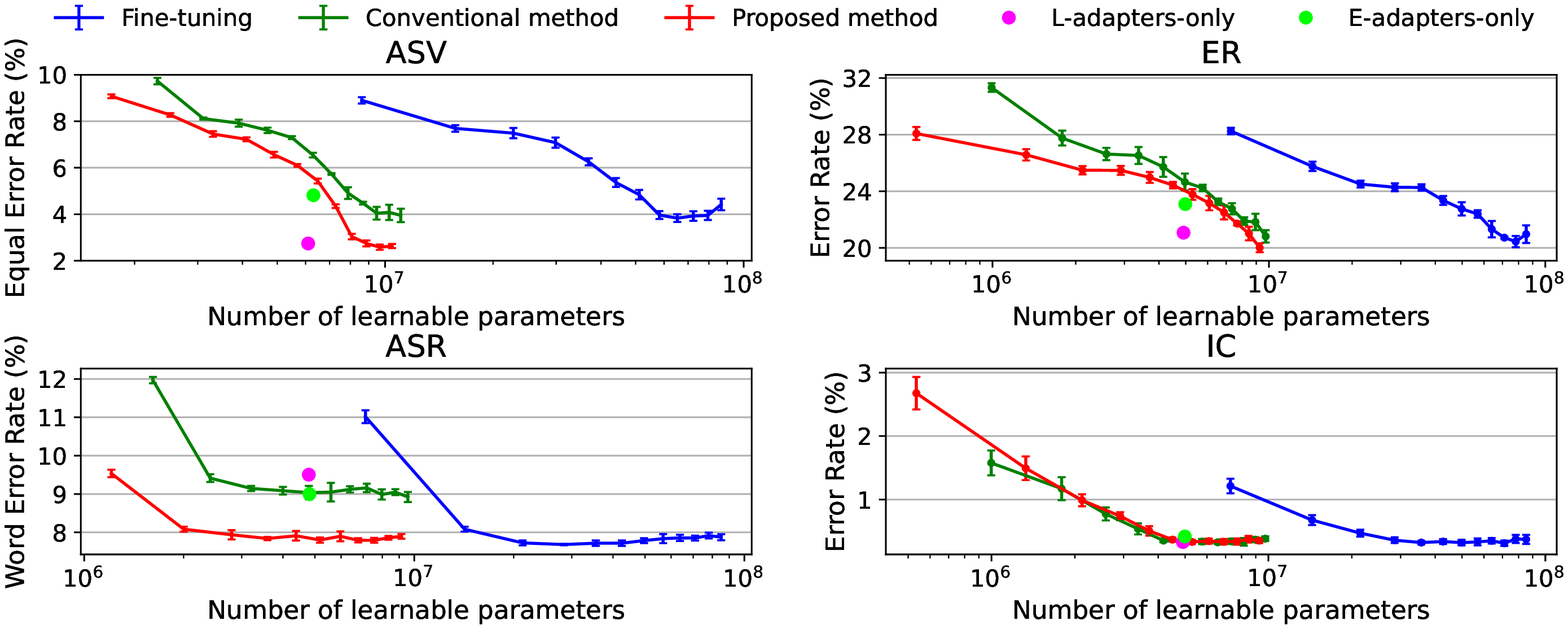}
  \vspace{-20pt}
\caption{
{\bf Experimental results for each task.} The x-axis represents the number of learnable parameters and the y-axis represents the error rate, both of which is better when they are smaller. Results only by L-adapters or E-adapters are shown for ablation.
}
\vspace{-10pt}
\label{fig:tuning-result}
\end{figure*}
}

\def\resultadapter{\begin{table*}[t]
    \centering
    \begin{tabular}{l|c|c|c|c|c|c}
    Types & \# Params & ASR WER & ASV EER & ER WA & IC ACC\\
    \hline
    Simple & 00.0M & $9.51 \pm 0.10$ & $2.74\pm 0.03 $ & $79.1\pm 0.37$ & $0.00 \pm 0.00$ \\
    only FC & 00.0M & $16.9 \pm 0.00$ & $4.36\pm 0.10$ & $75.8\pm 0.46$ & $0.00 \pm 0.00$ \\
    FC+Act & 00.0M & $10.2 \pm 0.00$ & $5.10\pm 0.10$ & $77.2\pm 0.51$ & $0.00 \pm 0.00$ \\
    FC+LN & 00.0M & $13.2 \pm 0.00$ & $\bm{2.70}\pm 0.04$ & $76.7 \pm 0.29$ & $0.00 \pm 0.00$ \\
    Act+LN & 00.0M & $21.8 \pm 0.00$ & $5.27\pm 0.06$ & $73.1 \pm 0.27$ & $0.00 \pm 0.00$ \\
    only LN & 00.0M & $22.0 \pm 0.00$ & $4.34\pm 0.06$ & $74.0 \pm 0.58$ & $0.00 \pm 0.00$ \\ \hline
    Skip & 00.0M & $0.00 \pm 0.00$ & $2.73$ & $80.0\pm 0.30$ & $0.00 \pm 0.00$\\
    \end{tabular}
    \caption{Caption}
    \label{tab:result}
\end{table*}
}

\def\resultadaptertmp{\begin{table}[t]
    \centering
    \begin{tabular}{l|c|c|c|c|c|c}
    Types & \# Params & ASR WER & ASV EER & ER WA & IC ACC\\
    \toprule
    Simple & 00.0M & $0.00 ^{\pm 0.00}$ & $2.70^{\pm 0.00} $ & $77.2^{\pm 0.00}$ & $0.00^{\pm 0.00}$ \\
    only FC & 00.0M & $16.9^{\pm 0.00}$ & $2.73^{\pm 0.00}$ & $75.8^{\pm 0.00}$ & $0.00^{\pm 0.00}$ \\
    FC+Act & 00.0M & $10.2^{\pm 0.00}$ & $5.03^{\pm 0.00}$ & $76.7^{\pm 0.00}$ & $0.00^{\pm 0.00}$ \\
    FC+LN & 00.0M & $13.2^{\pm 0.00}$ & $2.81^{\pm 0.00}$ & $76.6^{\pm 0.00}$ & $0.00^{\pm 0.00}$ \\
    Act+LN & 00.0M & $21.8^{\pm 0.00}$ & $5.34^{\pm 0.00}$ & $73.3^{\pm 0.00}$ & $0.00^{\pm 0.00}$ \\
    only LN & 00.0M & $22.0^{\pm 0.00}$ & $4.42^{\pm 0.00}$ & $74.8^{\pm 0.00}$ & $0.00^{\pm 0.00}$ \\ \hline
    Skip & 00.0M & $0.00^{\pm 0.00}$ & $2.73^{\pm 0.00}$ & $80.3^{\pm 0.00}$ & $0.00^{\pm 0.00}$\\
    \bottomrule
    \end{tabular}
    \caption{Caption}
    \label{tab:result}
\end{table}
}

\def\resultablation{\begin{table*}[t]
    \centering
    \begin{tabular}{l cc|cccc}
        Module & Proj size & \# Params & ASR WER& ASV EER& ER WA& IC ACC\\
        \toprule
        \multirow{2}{*}{FC} & 512 & 00.0M & $16.9 \pm 0.00$ & $4.36 \pm 0.10$ & $75.8 \pm 0.46$ & $0.00 \pm 0.00$ \\
        & 256 & 00.0M & $16.9 \pm 0.00$ & $4.36 \pm 0.10$ & $75.8 \pm 0.46$ & $0.00 \pm 0.00$ \\
        \multirow{2}{*}{FC+Act} & 512 & 00.0M & $10.2 \pm 0.00$ & $5.10 \pm 0.10$ & $77.2 \pm 0.51$ & $0.00 \pm 0.00$ \\
        & 256 & 00.0M & $10.2 \pm 0.00$ & $5.10 \pm 0.10$ & $77.2 \pm 0.51$ & $0.00 \pm 0.00$ \\
        \multirow{2}{*}{FC+LN} & 512 & 00.0M & $13.2 \pm 0.00$ & $2.70 \pm 0.04$ & $76.7 \pm 0.29$ & $0.00 \pm 0.00$ \\
        & 256 & 00.0M & $13.2 \pm 0.00$ & $2.70 \pm 0.04 $ & $76.7 \pm 0.29 $ & $0.00 \pm 0.00$ \\
        Act+LN & N/A & 00.0M & $21.8 \pm 0.00$ & $5.27 \pm 0.06 $ & $73.1 \pm 0.27$ & $0.00 \pm 0.00$ \\
        LN & N/A & 00.0M & $22.0 \pm 0.00 $ & $4.34 \pm 0.06 $ & $74.0 \pm 0.58$ & $0.00 \pm 0.00$ \\\hline
        \multirow{2}{*}{Simple} & 512 & 00.0M & $9.51 \pm 0.10$ & $2.74 \pm 0.03 $ & $79.1 \pm 0.37$ & $0.00 \pm 0.00$ \\
        & 256 & 00.0M & $9.51 \pm 0.10$ & $2.74 \pm 0.03 $ & $79.1\pm 0.37$ & $0.00 \pm 0.00$ \\
        \multirow{2}{*}{Skip} & 256 & 00.0M & $0.00 \pm 0.00$ & $2.73 \pm 0.00$ & $80.0 \pm 0.30$ & $0.00 \pm 0.00$\\
        & 128 & 00.0M & $0.00 \pm 0.00$ & $2.73 \pm 0.00$ & $80.0 \pm 0.30$ & $0.00 \pm 0.00$\\
        \bottomrule
    \end{tabular}
    \caption{Caption}
    \label{tab:result}
\end{table*}
}

\def\resultablationb{
\begin{table*}[t]
    \centering
    \begin{tabular}{l cc|cccc}
        Module & Proj size & \# Params & ASR WER& ASV EER& ER WA& IC ACC\\
        \toprule
        FC & 512 & 00.0M & $16.9 \pm 0.00$ & $4.36 \pm 0.10$ & $75.8 \pm 0.46$ & $0.00 \pm 0.00$ \\
        FC+Act & 512 & 00.0M & $10.2 \pm 0.00$ & $5.10 \pm 0.10$ & $77.2 \pm 0.51$ & $0.00 \pm 0.00$ \\
        FC+LN & 512 & 00.0M & $13.2 \pm 0.00$ & $2.70 \pm 0.04$ & $76.7 \pm 0.29$ & $0.00 \pm 0.00$ \\
        Act+LN & N/A & 00.0M & $21.8 \pm 0.00$ & $5.27 \pm 0.06 $ & $73.1 \pm 0.27$ & $0.00 \pm 0.00$ \\
        LN & N/A & 00.0M & $22.0 \pm 0.00 $ & $4.34 \pm 0.06 $ & $74.0 \pm 0.58$ & $0.00 \pm 0.00$ \\\hline
        \multirow{2}{*}{Simple} & 512 & 00.0M & $9.51 \pm 0.10$ & $2.74 \pm 0.03 $ & $79.1 \pm 0.37$ & $0.00 \pm 0.00$ \\
        & 256 & 00.0M & $9.51 \pm 0.10$ & $2.74 \pm 0.03 $ & $79.1\pm 0.37$ & $0.00 \pm 0.00$ \\
        \multirow{2}{*}{Skip} & 256 & 00.0M & $0.00 \pm 0.00$ & $2.73 \pm 0.00$ & $80.0 \pm 0.30$ & $0.00 \pm 0.00$\\
        & 128 & 00.0M & $0.00 \pm 0.00$ & $2.73 \pm 0.00$ & $80.0 \pm 0.30$ & $0.00 \pm 0.00$\\
        \bottomrule
    \end{tabular}
    \caption{Caption}
    \label{tab:result}
\end{table*}
}

\def\resultsmallablation{\begin{table*}[t]
    \centering
    \begin{tabular}{l cc|cccc}
        Module & Proj size & \# Params & ASR WER& ASV EER& ER WA& IC ACC\\
        \toprule
        \multirow{2}{*}{FC} & 512 & 00.0M & $16.9^{\pm 0.00}$ & $4.36^{\pm 0.10}$ & $75.8^{\pm 0.46}$ & $0.00^{\pm 0.00}$ \\
        & 256 & 00.0M & $16.9^{\pm 0.00}$ & $4.36^{\pm 0.10}$ & $75.8^{\pm 0.46}$ & $0.00^{\pm 0.00}$ \\
        \multirow{2}{*}{FC+Act} & 512 & 00.0M & $10.2^{\pm 0.00}$ & $5.10^{\pm 0.10}$ & $77.2^{\pm 0.51}$ & $0.00^{\pm 0.00}$ \\
        & 256 & 00.0M & $10.2^{\pm 0.00}$ & $5.10^{\pm 0.10}$ & $77.2^{\pm 0.51}$ & $0.00^{\pm 0.00}$ \\
        \multirow{2}{*}{FC+LN} & 512 & 00.0M & $13.2^{\pm 0.00}$ & $2.70^{\pm 0.04}$ & $76.7^{\pm 0.29}$ & $0.00^{\pm 0.00}$ \\
        & 256 & 00.0M & $13.2^{\pm 0.00}$ & $2.70^{\pm 0.04}$ & $76.7^{\pm 0.29}$ & $0.00^{\pm 0.00}$ \\
        Act+LN & N/A & 00.0M & $21.8^{\pm 0.00}$ & $5.27^{\pm 0.06}$ & $73.1^{\pm 0.27}$ & $0.00^{\pm 0.00}$ \\
        LN & N/A & 00.0M & $22.0^{\pm 0.00}$ & $4.34^{\pm 0.06}$ & $74.0^{\pm 0.58}$ & $0.00^{\pm 0.00}$ \\\hline
        \multirow{2}{*}{Simple} & 512 & 00.0M & $9.51^{\pm 0.10}$ & $2.74^{\pm 0.03} $ & $79.1^{\pm 0.37}$ & $0.00^{\pm 0.00}$ \\
        & 256 & 00.0M & $9.51^{\pm 0.10}$ & $2.74^{\pm 0.03} $ & $79.1^{\pm 0.37}$ & $0.00^{\pm 0.00}$ \\
        \multirow{2}{*}{Skip} & 256 & 00.0M & $0.00^{\pm 0.00}$ & $2.73^{\pm 0.00}$ & $80.0^{\pm 0.30}$ & $0.00^{\pm 0.00}$\\
        & 128 & 00.0M & $0.00^{\pm 0.00}$ & $2.73^{\pm 0.00}$ & $80.0^{\pm 0.30}$ & $0.00^{\pm 0.00}$\\
        \bottomrule
    \end{tabular}
    \caption{Caption}
    \label{tab:result}
\end{table*}
}

\def\result{\begin{table*}[t]
    \centering
    \begin{tabular}{l|c|c|c|c|c|c}
    Method & \# Params & ASR WER & ASV EER & ER WA & IC ACC\\
    \hline
    Weighted Sum & 00.0M & $0.00 \pm 0.00$ & $0.00 \pm 0.00$ & $0.00 \pm 0.00$ & $0.00 \pm 0.00$ \\
    Conventional Adapters & 00.0M & $0.00 \pm 0.00$ & $4.10 \pm 0.34$ & $77.6 \pm 0.35$ & $0.00 \pm 0.00$ \\
    \hline
    Fine-tuning all layers & 00.0M & $0.00 \pm 0.00$ & $4.42 \pm 0.25 $ & $79.2 \pm 0.77$ & $0.00 \pm 0.00$ \\
    L-Adapters alone (Ours) & 00.0M & $9.51 \pm 0.10$ & $\bm{2.70} \pm 0.04 $ & $\bm{80.0} \pm 0.30$ & $0.00 \pm 0.00$ \\
    Proposed Adapters (Ours)& 00.0M & $0.00 \pm 0.00$ & $0.00 \pm 0.00$ & $0.00 \pm 0.00$ & $0.00 \pm 0.00$ \\
    \end{tabular}
    \caption{Caption}
    \label{tab:result}
\end{table*}
}

\begin{abstract}
Fine-tuning of self-supervised models is a powerful transfer learning method in a variety of fields, including speech processing, since it can utilize generic feature representations obtained from large amounts of unlabeled data. Fine-tuning, however, requires a new parameter set for each downstream task, which is parameter inefficient. 
Adapter architecture is proposed to partially solve this issue by inserting lightweight learnable modules into a frozen pre-trained model. 
However, existing adapter architectures fail to adaptively leverage low- to high-level features stored in different layers, which is necessary 
for solving various kinds of speech processing tasks.
Thus, we propose a new adapter architecture to acquire feature representations more flexibly for various speech tasks. In experiments, we applied this adapter to WavLM on four speech tasks. It performed on par or better than na\"{i}ve fine-tuning, with only 11\% of learnable parameters. It also outperformed an existing adapter architecture.
Our implementation code is available at 
\url{https://github.com/sinhat98/adapter-wavlm}
\end{abstract}

\section{Introduction}
Self-supervised learning with large-scale unlabeled datasets has become one of the most promising approaches for learning generic speech representations.
For example, recent models such as wav2vec 2.0 \cite{wav2vec2}, HuBERT~\cite{hubert}, and WavLM~\cite{wavlm} have successfully extracted task-independent representations from speech data.

However, how to efficiently utilize such rich speech representations for various downstream tasks is an open issue. Particularly, we consider the case where multiple downstream tasks are required to be learned independently. To build a model that performs well on each task, a potential approach is to fine-tune a pre-trained model with the label of the target downstream task \cite{fine-tuning, finetune-sv, finetune-er}. Fine-tuning performs well, but parameter efficiency is poor because it requires a new complete set of parameters for each task. Another approach is multi-task learning, which shares network parameters across interrelated tasks to improve performance and efficiency \cite{multitask1, multitask2}. However, learning less relevant tasks simultaneously causes degradation of performance \cite{multitask3}. In addition, it is impossible to learn tasks independently in multi-task learning.
Adapters can be one of the solutions for transfer learning that are parameter efficient yet achieve good performance. They are lightweight modules that can be inserted into some intermediate layers of a frozen pre-trained model and require fewer parameters because the frozen parameters are shared among all downstream tasks.
Adapters are proposed for BERT \cite{bert} in natural language processing
as an efficient method of transfer learning \cite{nlp-adapter}, and 
they are shown to be effective for  multi-lingual speech recognition \cite{rnnt-adapter, ctcattn-adapter} and speech translation~\cite{sl-adapter} as well.
They are also applied to
 wav2vec~2.0 by Tomas {\it et al.} \cite{speech-adapter}, which achieves a lightweight transfer learning that performs on par with fine-tuning in automatic speech recognition. 
 
Applying these adapters directly to solve multiple speech processing tasks, however, is not trivial. The existing adapter architectures rely on the high similarity between the self-supervised pre-training task and the target task; thus, they may not work well,
for example in speaker-related tasks, because they cannot adaptively extract speech representations in different layers of the pre-trained model.
Designing effective adapter architecture for various speech tasks is challenging because the features required to solve a task can be different from task to task. Chen {\it et al.} \cite{wavlm} have found that different layers are important for different tasks in WavLM.
It is also reported that various acoustic and linguistic features tend to be encoded in different layers in wav2vec~2.0~\cite{analysis1, analysis2}.
These studies motivated us to explore a novel adapter architecture.

In this work, we propose a novel adapter architecture that is effective for various speech tasks.
The proposed architecture inserts small learnable modules between each intermediate layer and the top layer with learnable weighting coefficients.
This enables the model to automatically find intermediate layers from which features should be extracted in a task-dependent manner.
In experiments, we evaluate the efficiency and the effectiveness of the proposed architecture in four downstream tasks, namely, automatic speech recognition (ASR), automatic speaker verification (ASV), emotion recognition (ER), and intent classification (IC).
With the WavLM backbone, our method achieved performance on par with fine-tuning while using 89\% fewer learnable parameters. We also compared our method with the conventional method in \cite{speech-adapter}, and showed that our method is more effective.
Further, we visualized the weight coefficients for each layer to explain why our method works well.

\section{Adapter Architecture}

The proposed adapter architecture incorporates two types of adapters, namely Layer adapters (L-adapters) and Encoder adapters (E-adapters), into a frozen backbone.
The L-adapters bridge each intermediate layer and the top layer, as shown in Figure~\ref{figlayeradapter}a.
They help the model to quickly adapt speech representations to various downstream tasks, and also to reduce dependency on the initialization of adapter parameters.
The E-adapters are inserted within each encoder layer in a similar way to in previous work \cite{nlp-adapter, speech-adapter}, as shown in Figure~\ref{figlayeradapter}b.
In contrast to the previous work, our architecture does not have adapters after the multi-head self-attention (MHSA) modules, but has L-adapters instead.

\noindent {\bf Overview.}
Let $F_{\theta}$ be a backbone model that maps an audio input $\bm{x}$ into a speech representation $\bm{y}$ as $\bm{y} = F_{\theta}(\bm{x})$ where $\theta$ is a set of backbone parameters.
Given a pre-trained parameter $\hat{\theta}$, the proposed adapter architecture freezes $\hat{\theta}$ and incorporates lightweight learnable adapters and a downstream head for fine-tuning.
As shown in Figure~\ref{figlayeradapter}a, we use WavLM for $F_{\theta}$, which consists of a CNN for pre-processing and $L$ encoder layers. Note that in the figure, the learnable modules are marked in red or yellow, and the frozen modules are marked in gray. The layer normalization modules of each encoder layer are learnable following the previous work in \cite{speech-adapter}.

\noindent {\bf L-adapters.}
To utilize intermediate representations from the early phases of fine-tuning, L-adapters make paths from each encoder layer to the downstream head.
Let $\bm{h}_{l}$ be the output of the $l$-th encoder layer.
The L-adapter $f_{l}$ is applied to it to obtain adapted representations as $\bm{a}_{l} = f_{l}(\bm{h}_{l})$ for $l = 1, 2, \cdots, L$.
The weighted sum of the adapted representations $\bm{h}^{*} = \sum_{l=1}^{L} w_{l} \bm{a}_{l}$ is then fed into the downstream head, where $w_{l}$ are learnable weights.
Each L-adapter consists of a fully connected (FC) layer followed by non-linear activation (Act) and layer normalization (LN), as shown in Figure~\ref{figlayeradapter}c. For non-linear activation, ReLU \cite{relu} is used for ASV and ER, and GELU \cite{gelu} is used for ASR and IC.
The embedding size is 512 for all L-adapters.

\noindent {\bf E-adapters.}
To obtain fine-grained representations via fine-tuning, E-adapters are incorporated into encoder layers.
Assuming that each encoder layer has a fully connected feedforward network (FFN), the E-adapter is inserted just after it as shown in Figure~\ref{figlayeradapter}b.
Each E-adapter consists of a two-layer MLP with LN and a skip connection (Figure~\ref{figlayeradapter}d).
For non-linear activation, the same function as for the L-adapters is used. The embedding size is 256 for all E-adapters.

\figlayeradapter
\noindent {\bf Downstream heads.}
The configuration of the downstream head is shown in Figure~\ref{figlayeradapter}e.
For ASR, CTC loss \cite{ctc} is applied over an FC layer.
For the other tasks, cross-entropy loss is minimized over two FC layers with average time pooling (ATPool) in between.
Note that speaker embeddings are extracted from ATPool in the evaluation phase of ASV.

\figresults
\section{Experiments}

\subsection{Experimental Settings}

We demonstrate the effectiveness of the proposed method on four downstream tasks: ASV, ER, ASR, and IC.
They belong to the four different aspects of speech \cite{superb}: speaker, paralinguistics, content, and semantics, respectively.
In the following, we provide the details of the experimental settings.

\noindent {\bf Backbone architecture.}
The WavLM Base model \cite{wavlm} is used as a backbone. It consists of 12 transformer encoder layers with gated relative position bias, 768-dimensional hidden states, and 8 attention heads,
which results in 94.7M parameters.
We used the pre-trained model, wavlm-base-plus\footnote{https://huggingface.co/microsoft/wavlm-base-plus}, which is obtained from self-supervised learning using the masked speech prediction task on 94k hours of public audio material. 

\noindent {\bf Conventional adapter architecture.}
We compare our architecture with the state-of-the-art adapter architecture in \cite{speech-adapter}, which has two adapter modules at each encoder layer, one after the MHSA and another after the feedforward module. Each adapter module consists of two fully connected layers with a GELU activation in between followed by layer normalization and a skip connection. 

\noindent {\bf Training methods.}
We run experiments on five training methods as follows. 
(i)~Fine-tuning the top $l$ layers for $l=1,2, \cdots ,12$.
(ii)~Conventional method: Adapters are inserted after MHSA and feedforward modules in the top $l$ layers of for $l=1,2,\cdots,12$. (iii)~Proposed method: L-adapters are attached to the top $k$ layers for $k=1,2,\cdots,12$ and E-adapters are inserted in the $l$ layers from the second layer from the top for $l=1,2,\cdots, 11$. (iv)~L-adapters-only: L-adapters are attached to all layers without E-adapters. (v)~E-adapters-only: E-adapters are inserted into all layers without L-adapters. 
We compare baseline methods (i), (ii) with our methods unique to this work (iii), (iv), (v).

\noindent {\bf Datasets and evaluation metrics.}
For ASV, the VoxCeleb1 dataset \cite{voxceleb} is used for evaluation. The dev subset, which consists of 148,642 utterances from 1,251 speakers and is approximately 351 total hours in length, is used for training. The cleaned original test set which consists of 37,611 trials over 40 speakers is used for testing. The evaluation metric is the equal error rate (EER). 
The trial scores are computed by the cosine similarity between speaker embeddings and are normalized using the adaptive s-norm \cite{snorm1, snorm2}. 

For ER, the IEMOCAP \cite{iemocap} dataset is used for evaluation.
It contains approximately 12 hours of audio data with scripted and improvised dialogues by 10 speakers.
We performed five-folds cross-validation, using  the emotion labels ``neutral'', ``happy'', ``sad'', and ``angry''. Note that, following the previous work in \cite{erpaper}, ``excited'' is merged into ``happy''. The evaluation metric is the error rate, which is defined as $1.0 - \mathrm{weighted\_accuracy}$. The weighted accuracy is the average of the classification accuracy per class.

For ASR, the LibriLight dataset \cite{librilight} (10-hour supervised subset) is used for training. This dataset is a collection of spoken English audio derived from open-source audio books.
The standard LibriSpeech dev set \cite{librispeech} is used for testing, and the evaluation metric is word error rate (WER).

\figminipage

For IC, the Fluent Speech Commands \cite{speechcommand} dataset is used for evaluation.
Fluent Speech Commands is a dataset of 30,043 English audio material with 77 speakers, approximately 19 hours in total, each labeled with ``action", ``object", and ``location" slots.
The train subset is used for training and the test subset is used for testing.
The evaluation metric is the error rate, which is defined as $1.0 - \mathrm{classification\_accuracy}$.

\noindent {\bf Overall training settings}
We trained for 18k, 2.8k, 35k, and 10k steps for ASV, ER, ASR, and IC, respectively.
The optimizer is RAdam \cite{radam} only for the case of fine-tuning in ASV, and Adam \cite{adam} for the others. 
We chose the best learning rates from \{1e-3, 5e-4, 1e-4, 5e-5, 1e-5\} for each architecture.

\subsection{Experimental Results}

{\noindent {\bf Main results.}}
Figure~\ref{fig:tuning-result} shows the performance comparison. 
The performance improvement figures below are comparisons at the right end of each curve, i.e., when all layers are used.
In ASV and ER, the reduction in absolute error compared to fine-tuning is $1.79\%$ for ASV and $0.97\%$ for ER. In ASR, the performance is comparable to that of fine-tuning, and despite the significant reduction in the number of task-specific parameters, the degradation for WER is only $0.03\%$. In IC, the proposed method shows a slightly better result which is a $0.02\%$ reduction in the error rate compared with fine-tuning. In addition, the proposed method performs better than the conventional method in all tasks.
Therefore, the proposed method performs well despite the small number of trained task-specific parameters over various speech processing tasks. This is a great benefit in performing multiple speech processing tasks in terms of parameter efficiency and scalability.
We also run experiments by varying the number of layers to fine-tune and to insert adapters to find cases where a smaller number of trained parameters would perform well. 
Fig.~\ref{fig:tuning-result} shows all the results of fine-tuning, conventional and proposed methods, as well as L-adapters-only and E-adapters-only for ablation.
In ASR and IC, the error curve of fine-tuning decreases quickly; only upper-layer adjustments are sufficient for these tasks. In contrast, the curves for ASV and ER decrease slowly until lower-layer adjustments are included.
This tendency that ASV and ER require lower-layer adjustment is also consistent in the conventional and the proposed methods.
L-adapters-only works well in ASV and ER, since it leverages outputs of all layers; it is a good model in terms of parameter efficiency and performance for these tasks. 
In ASR and IC, the proposed method with adapters applied to the upper 6 or 7 layers performed the best.

{\noindent {\bf Robustness to the learning rate and the initialization.}}
Through the learning rate search, the conventional method and E-adapters-only experience cases where the loss does not converge with high learning rates, which leads to performance degradation. 
For high learning rates, the initialization of adapters has a considerable impact on training in the conventional method and E-adapters-only. Specifically, the convergence of losses and the error rates vary by the different seeds.
In contrast, the losses converge even at high learning rates with the proposed method because L-adapters adapt each layer to the downstream head.
Therefore, tuning of the proposed adapter is easier than tuning the conventional one.

{\noindent {\bf L-adapter configurations.}}
To investigate the most effective configuration for L-adapters, we conducted experiments with eight configurations. The results are reported in Table~\ref{result1}.
Note that Weight is the weighted sum of the encoder outputs. Additionally, Base and Skip are the configurations in Figure~\ref{figlayeradapter}c and ~\ref{figlayeradapter}d, respectively.
We see that the Base configuration consistently works well. This shows that the combination of FC, Act, and LN is necessary for most downstream tasks. Act can be omitted for ASV and IC but this does not improve the parameter efficiency.
For only ASR, FC+Act performs slightly better than Skip. This suggests that the best configuration could have been different from task to task. However, our conclusion here is that the Base configuration is the most reasonable.
Note that, for E-adapters, the Skip configuration was the best. This is the same conclusion as in most previous work \cite{nlp-adapter, speech-adapter}.

{\noindent {\bf Layer weight analysis.}} 
To confirm the contribution of L-adapters and E-adapters, Figure~\ref{fig:layer-weight-analysis} visualizes the learned weight coefficients for each layer obtained in L-adapters-only and the proposed method.
In ASR, the weights of features obtained from the upper layers are larger, while in ASV the weights of the lower layers tend to be larger. 
In ER and IC, all layers are leveraged almost equally.
These tendencies are the factors contributing to the superiority of L-adapters. Fig.~\ref{fig:layer-weight-analysis} also shows how the weights change by the E-adapters adaptation. Only in ASR, the distribution of weights changes and the weight of the final layer becomes large due to the E-adapters. 
In non-ASR tasks, such changes are not seen.  
This indicates that adapter architectures that connect modules in series such as the conventional architecture would not work well for transferring in non-ASR tasks.

\section{Conclusion}
The conventional adapter architecture is not ideal for transferring effective features in non-ASR tasks. This is because transformer-based self-supervised speech models have different speech representations in different layers. L-adapters can obtain effective features by selecting those that are important for downstream tasks among the various speech representations in different layers. E-adapters can improve the adaptability of the overall network. We have combined L-adapters and E-adapters to increase the adaptability to downstream tasks and achieved a general-purpose transfer method. 

In this paper, we proposed a transfer learning method for multiple speech processing tasks. 
It is found that adapting the output of each layer downstream is effective for applying the pre-trained model to a wide range of tasks. 
In future work, designing adapter architectures that can obtain more generic features more efficiently and extending the proposed method
to multi-lingual and multi-modal scenarios could be done.

\small
\bibliography{main.bib}

\begin{thebibliography}{10}

\bibitem{wav2vec2}
A.~Beavski {\it et al}.,
\newblock ``{wav2vec 2.0: A Framework for Self-Supervised Learning of Speech
  Representations},''
\newblock {\em Advances in Neural Information Processing Systems}, vol. 33, pp.
  12449--12460, 2020.

\bibitem{hubert}
W.N.~Hsu {\it et al.},
\newblock ``{HuBERT: Self-Supervised Speech Representation Learning by Masked
  Prediction of Hidden Units},''
\newblock {\em IEEE/ACM Transactions on Audio, Speech, and Language
  Processing}, vol. 29, pp. 3451--3460, 2021.

\bibitem{wavlm}
S.~Chen {\it et al.},
\newblock ``{WavLM: Large-Scale Self-Supervised Pre-Training for Full Stack
  Speech Processing},''
\newblock {\em IEEE Journal of Selected Topics in Signal Processing}, 2022.

\bibitem{fine-tuning}
Y.~Wang {\it et al.},
\newblock ``{A Fine-tuned Wav2vec 2.0/HuBERT Benchmark For Speech Emotion
  Recognition, Speaker Verification and Spoken Language Understanding},''
\newblock {\em arXiv preprint arXiv:2111.02735}, 2021.

\bibitem{finetune-sv}
N.~Vaessen {\it et al.},
\newblock ``{Fine-Tuning Wav2Vec2 for Speaker Recognition},''
\newblock in {\em Proc. ICASSP}, 2022, pp. 7967--7971.

\bibitem{finetune-er}
L.~Pepino {\it et al.},
\newblock ``{Emotion Recognition from Speech Using wav2vec 2.0 Embeddings},''
\newblock in {\em Proc. Interspeech}, 2021, pp. 3400--3404.

\bibitem{multitask1}
X.~Cai {\it et al.},
\newblock ``{Speech Emotion Recognition with Multi-Task Learning},''
\newblock in {\em Proc. Interspeech}, 2021, pp. 4508--4512.

\bibitem{multitask2}
S.~Hussain {\it et al.},
\newblock ``{Multi-Task Voice Activated Framework Using Self-Supervised
  Learning},''
\newblock in {\em Proc. ICASSP}, 2022, pp. 6137--6141.

\bibitem{multitask3}
Y.C.~Chen {\it et al.},
\newblock ``{Speech Representation Learning Through Self-supervised Pretraining
  And Multi-task Finetuning},''
\newblock in {\em Proc. AAAI Workshop on Self-supervised Learning for Audio and
  Speech Processing}, 2022.

\bibitem{bert}
J.~Devlin {\it et al.},
\newblock ``{BERT: Pre-training of Deep Bidirectional Transformers for Language
  Understanding},''
\newblock in {\em Proc. NAACL}, 2019.

\bibitem{nlp-adapter}
N.~Houlsby {\it et al.},
\newblock ``{Parameter-Efficient Transfer Learning for NLP},''
\newblock in {\em Proc. ICML}, 2019, pp. 2790--2799.

\bibitem{rnnt-adapter}
A.~Kannan {\it et al.},
\newblock ``{Large-Scale Multilingual Speech Recognition with a Streaming
  End-to-End Model},''
\newblock in {\em Proc. Interspeech}, 2019.

\bibitem{ctcattn-adapter}
G.I.~Winata {\it et al.},
\newblock ``{Adapt-and-Adjust: Overcoming the Long-Tail Problem of Multilingual
  Speech Recognition},''
\newblock in {\em Proc. Interspeech}, 2021.

\bibitem{sl-adapter}
H.~Le {\it et al.},
\newblock ``{Lightweight Adapter Tuning for Multilingual Speech Translation},''
\newblock in {\em Proc. ACL}, 2021.

\bibitem{speech-adapter}
B.~Thomas {\it et al.},
\newblock ``{Efficient Adapter Transfer of Self-Supervised Speech Models for
  Automatic Speech Recognition},''
\newblock in {\em Proc. ICASSP}, 2022, pp. 7102--7106.

\bibitem{analysis1}
A.~Pasad {\it et al.},
\newblock ``{Layer-wise Analysis of a Self-supervised Speech Representation
  Model},''
\newblock in {\em Proc. 2021 IEEE Automatic Speech Recognition and
  Understanding Workshop (ASRU)}, 2021, pp. 914--921.

\bibitem{analysis2}
J.~Shah {\it et al.},
\newblock ``{What all do audio transformer models hear? Probing Acoustic
  Representations for Language Delivery and its Structure},''
\newblock {\em arXiv preprint arXiv:2101.00387}, 2021.

\bibitem{relu}
A.F. Agarap,
\newblock ``{Deep Learning using Rectified Linear Units (ReLU)},''
\newblock {\em arXiv preprint arXiv:1803.08375}, 2018.

\bibitem{gelu}
D.~Hendrycks {\it et al.},
\newblock ``{Gaussian Error Linear Units (GELUs)},''
\newblock {\em arXiv preprint arXiv:1606.08415}, 2016.

\bibitem{ctc}
A.~Graves {\it et al.},
\newblock ``{Connectionist Temporal Classification: Labelling Unsegmented
  Sequence Data with Recurrent Neural Networks},''
\newblock in {\em Proc. ICML}, 2006, pp. 369--376.

\bibitem{superb}
S.~Yang {\it et al.},
\newblock ``{SUPERB: Speech Processing Universal PERformance Benchmark},''
\newblock in {\em Proc. Interspeech 2021}, 2021, pp. 1194--1198.

\bibitem{voxceleb}
A.~Nagrani {\it et al.},
\newblock ``{VoxCeleb: A Large-Scale Speaker Identification Dataset},''
\newblock in {\em Proc. Interspeech}, 2017, pp. 2616--2620.

\bibitem{snorm1}
Z.N.~Karam {\it et al.},
\newblock ``Towards reduced false-alarms using cohorts,''
\newblock in {\em Proc. ICASSP}, 2011, pp. 4512--4515.

\bibitem{snorm2}
S.~Cumani {\it et al.},
\newblock ``{Comparison of Speaker Recognition Approaches for Real
  Applications.},''
\newblock in {\em Proc. Interspeech}, 2011, pp. 2365--2368.

\bibitem{iemocap}
C.~Busso {\it et al.},
\newblock ``{IEMOCAP: Interactive emotional dyadic motion capture database},''
\newblock {\em Language resources and evaluation}, vol. 42, no. 4, pp.
  335--359, 2008.

\bibitem{erpaper}
H.M.~Fayek {\it et al.},
\newblock ``Evaluating deep learning architectures for speech emotion
  recognition,''
\newblock {\em Neural Networks}, vol. 92, pp. 60--68, 2017.

\bibitem{librilight}
J.~Kahn {\it et al.},
\newblock ``{Libri-Light: A Benchmark for ASR with Limited or No
  Supervision},''
\newblock in {\em Proc. ICASSP}, 2020, pp. 7669--7673.

\bibitem{librispeech}
V.~Panayotov {\it et al.},
\newblock ``{Librispeech: An ASR corpus based on public domain audio books},''
\newblock in {\em Proc. ICASSP}, 2015, pp. 5206--5210.

\bibitem{speechcommand}
L.~Lugosch {\it et al.},
\newblock ``{Speech Model Pre-training for End-to-End Spoken Language
  Understanding},''
\newblock in {\em Proc. Interspeech}, 2019.

\bibitem{radam}
L.~Liu {\it et al.},
\newblock ``{On the Variance of the Adaptive Learning Rate and Beyond},''
\newblock {\em arXiv preprint arXiv:1908.03265}, 2019.

\bibitem{adam}
D.P.~Kingma {\it et~al.},
\newblock ``{Adam: A Method for Stochastic Optimization},''
\newblock {\em arXiv preprint arXiv:1412.6980}, 2014.

\end{thebibliography}
\bibliographystyle{temp.bst}

\end{document}